# OpenHI2 – Open source histopathological image platform

Pargorn Puttapirat[1,2], Haichuan Zhang[1,2], Jingyi Deng[2,3], Yuxin Dong[1,2], Jiangbo Shi[1,2], Hongyu He[4], Zeyu Gao[1,2], Chunbao Wang[4], Xiangrong Zhang[5], Chen Li[1,2]*
[1] Shaanxi Province Key Laboratory of Satellite and Terrestrial Network Tech. R&D，Xi'an Jiaotong University, Xi'an, Shaanxi 710049, China
[2] School of Electronic and Information Engineering，Xi'an Jiaotong University, Xi'an, Shaanxi 710049, China
[3] Department of Software Engineering, Xi'an Jiaotong University, Xi'an, Shaanxi 710049, China
[4] Department of Power System and Automation, School of Electrical Engineering, Xi'an Jiaotong University, Xi'an, Shaanxi 710049, China
[5] Department of Pathology, the First Affiliated Hospital of Xi'an Jiaotong University, Xi'an, Shaanxi 710061, China
[6] Institute of Intelligent Information Processing, Xidian University, Xi'an, Shaanxi 710071, China
Email: {pargorn, zhanghaichuan, misscc320, dongyuxin, shijiangbo}@stu.xjtu.edu.cn, nostalgia1229@gmail.com, gzy4119105156@stu.xjtu.edu.cn, bingliziliao2012@163.com, xrzhang@mail.xidian.edu.cn, cli@xjtu.edu.cn*

*Abstract*— Transition from conventional to digital pathology requires a new category of biomedical informatic infrastructure which could facilitate delicate pathological routine. Pathological diagnoses are sensitive to many external factors and is known to be subjective. Only systems that can meet strict requirements in pathology would be able to run along pathological routines and eventually digitized the study area, and the developed platform should comply with existing pathological routines and international standards. Currently, there are a number of available software tools which can perform histopathological tasks including virtual slide viewing, annotating, and basic image analysis, however, none of them can serve as a digital platform for pathology. Here we describe OpenHI2, an enhanced version Open Histopathological Image platform which is capable of supporting all basic pathological tasks and file formats; ready to be deployed in medical institutions on a standard server environment or cloud computing infrastructure. In this paper, we also describe the development decisions for the platform and propose solutions to overcome technical challenges so that OpenHI2 could be used as a platform for histopathological images. Further addition can be made to the platform since each component is modularized and fully documented. OpenHI2 is free, open-source, and available at https://gitlab.com/BioAI/OpenHI.

*Keywords—OpenHI, digital pathology, histopathological image, analysis, open source software*

I. INTRODUCTION

Rapid developments in artificial intelligence (machine learning) techniques could directly benefit pathology by alleviating laborious tasks formerly performed by pathologists. However, difficulties in transition to digital pathology have be obstructing advances in this area. An open platform which is has functions to support pathology routine, extendable, and standardized is needed to facilitate the transition from conventional to digital pathology. In this paper, we present the updated version of OpenHI – Open Histopathological Image [1], [2] which has evolved to be a more general platform for digital pathology rather than being a web-based annotation framework. OpenHI2 provides functions that facilitate basic pathological tasks including slide viewing, diagnostic region selection, digital magnification based on conventional magnification, semantic annotation, and standardized clinical record viewing. As well as the first version, OpenHI2 is developed as a web-based application thus it adopts server-client scheme similar to the picture archiving and communication system (PACS) widely used in radiology. The proposed platform has incorporated the needs from pathologists and data scientists. It can potentially be used as a part of biomedical informatics infrastructure for pathology and promote machine learning based analysis on digital histopathological slides.

Continuous efforts toward digital pathology roughly started in 1999 with the idea of slide digitization proposed by Wetzel and Gilbertson [3]. Followed by large-scale project that collected digital slides including TCGA in 2005 [4] and GTEx in 2013 [5]. Later, digital slides are also known as whole-slide images (WSIs) which are produced by whole-slide scanners developed by different companies resulting in independent file formats. Vendor neutral digital slide reading library and bioimage platform such as OpenSlide [6] and BioFormat [7] are proposed and they are important parts of digital pathology. Since the introduction of OpenSlide, a number of stand-alone software such as QuPath [8], Icy [9], and SlideRunner [10] and web-based framework such as Cytomine [11] and Cancer Digital Slide Archive Project [12] which are potentially useful in pathology have emerged. The mentioned works have proven that it is possible to view digital slides through computer systems. The next phase of transition from conventional to digital pathology is for the software to serve as a biomedical informatic platform. Advantage of such digitized platform is that it could systematically capture pathological routines and medical decisions made by clinicians and pathologists. Thus, captured information may be used to establish statistical machine learning based systems, especially since current proposed models lacks large-scale annotated datasets to train upon.

In principle, pathological tasks could be categorized into three group: slide viewing, analyzing, and storage. Different functions of OpenHI2 will be discussed in this paper. In slide viewing, there are many WSI viewer readily available and there are very few studies [13] which investigate the effect that virtual slides have on the final diagnosis. In addition, the effect may be specific to different tumor type as well. Concerns and proposed solutions regarding displaying virtual slides have been made by researchers [14] and US Food and Drug Administration (US FDA) [15], [16]. All existing software and framework that currently support WSIs did not have a clear solution regarding this issue. Analyzing WSIs require subtle modifications to be made to general image analysis tools. For instance, processing on the entire WSI may be unnecessary since diagnosis can be made, conventionally, by examinations on some areas. In that case, attentions mechanisms are being utilized. Finally, the nature of being huge in file size of WSIs pose some challenges in storing and accessing the repository. Integrations of an efficient library to handle WSI files are demonstrated in OpenHI2 through its web-based framework.

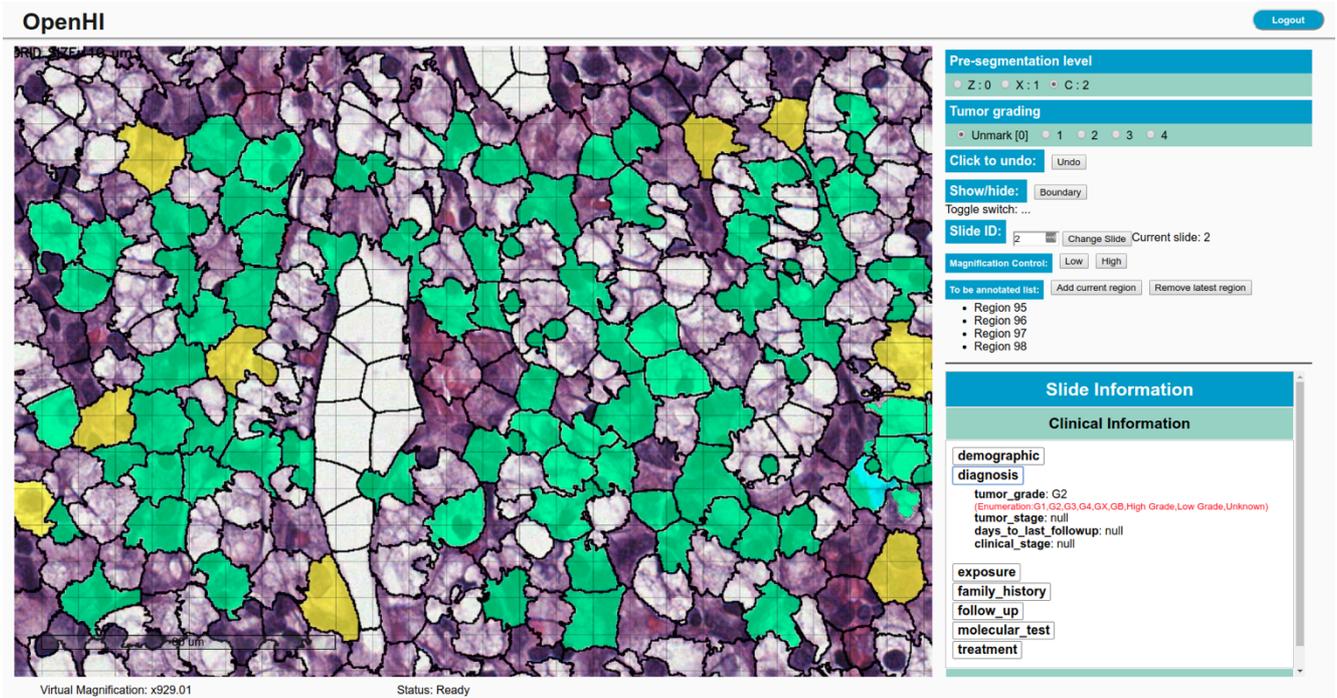

Fig. 1. Web-based graphic user interface of OpenHI2 including the viewer (left), annotation control module (top-right), and slide information (bottom-right). In this figure, the viewer is fixed at 1200-by-900 pixel and the entire GUI is displayed on 1920-by-1080 pixel display.

## II. RELATED WORK

Problems regarding virtual slide viewing especially in WSIs where no conventional light microscope magnification mechanisms are being used: when a CCD camera is attached to the microscope via eyepiece lenses, is roughly pointed out by the US FDA as image pathway issue. With the mentioned concept, manufacturers of WSI systems should be able to describe how light originate from the light source, travel through the sample, being digitized, and viewed as pixels on computer displays. In other words, one should know what a single pixel on a computer display represents. In 2013, T. Sellaro et al. [14] describe a method to trace back the origin of pixels in WSIs and mention the differences between conventional and digital slides. However, it is difficult to control the different models of display on a web-based or cross-platform application thus it is hard to implement this method. Therefore, this remains a current issue and will be discussed in this paper.

Better medical image processing and classification algorithms in histopathology have emerged rapidly over the years [17]–[19]. Many of them only take selected regions provided by public datasets derived from competitions [20], [21], thus, information at the macro level is being discarded. Meanwhile, [22] is a preliminary evidence proof that visual information at the macro or whole-slide level do reflect the patient's outcome. To promote further development in this area, standardized annotation framework which support large-scale collaborative annotation is needed. The first version of OpenHI [1] can be used as an annotation tool for this occasion. Furthermore, training newer models can be done easily by utilizing basic image processing functions in the image processing module in OpenHI2.

Currently, there are only a few libraries and file formats that could handle extremely large histopathological images including OpenSlide—a vendor-neutral reading library [6], BioFormat which is a part of open microscope environment project [7], and our in-house library—libMI—along with a compressive file format—MMSI—which could efficiently store WSIs. libMI is an extension of OpenSlide library, therefore it can read previously support WSI file formats which cover all major formats produced by WSI scanners. Previous attempts on establishing digital slide server are successful [10]–[12], relying on web-based image viewer such as OpenSeadragon project [23] which make them good for digital slide viewing and annotation. However, providing further functions and support for machine learning algorithms could be limiting.

## III. MATERIALS AND METHODS

The proposed platform was not aimed to replace the whole pathological process. It only facilitates pathological routines once the digital slides are acquired by whole-slide scanners. As mentioned earlier, pathological tasks involving histopathological images can be categorized into three groups, thus we will discuss about requirements, then our proposed methods to develop the platform so that it can serve as platform for digital pathology.

The pathological and clinical requirements discussed here derived from two sources: experienced pathologists and selected international pathological diagnosis standards. The pathologists would provide insights to how diagnoses are routinely made in pathological laboratories, while international standards would ensure the quality of the platform that it can comply with such standard.

*A. The platform*

OpenHI2 is developed based on the previous version which is a web-based annotation framework for histopathological images [1]. The infrastructure of the software is flexible, extensible, and easy to use via the GUI already. Thus, we are keeping similar development scheme in this version as well. The new developments in this version includes viewing configuration tools which complies with our newly proposed method to ensure the quality of WSIs displayed on user's computer screen, a module to select diagnostic regions which facilitate a more harmonize image annotation between multiple annotators, basic statistical machine learning based module, and a module for inter-rater reliability tests.

*1) Virtual slide viewing*

All WSI-supported tools allow users to freely zoom and pan around the WSI. While to freely zoom is newly introduced in digital slides, it came with an issue of "not knowing the magnification" relative to magnification factors which is fixed and can be easily calculated in general optical microscopes: objective multiplied by eye piece lenses magnification. The lack of this reference makes other platforms incompatible with histopathological grading standards that rely on fixed magnification in optical microscope such as the renal cell carcinoma grading system maintained by The International Society of Urological Pathology (ISUP) which specifically require pathologist to examine the slide using 100x and 400x magnifications.

TABLE I. SAMPLE CALCULATION OF REQUIRED WSI PIXEL RESOLUTION/SIZE BASED ON RAYLEIGH RESOLUTION LIMIT.

| Magnification | 100x | 400x |
|---|---|---|
| Typical N.A. | 0.25 | 0.65 |
| Resolving power (wavelength= 0.56) [micron] | 1.456 | 0.56 |
| Required pixel size [micron] | 0.728 | 0.280 |
| Down sampling factor | 2.9 ≈ 3 | 1.12 ≈ 1 |

In response to this issue, we propose a new approach for displaying the image on the viewer, focusing on the down sampling factor. The approach leverage the fact that objective lenses in the microscope has a resolution limits which are corresponding to the magnification and can be calculated using Rayleigh resolution limit [24] by supplying the numerical aperture (N.A.) number of the objective lenses. After that, the resolution should be double in digital image according to the Nyquist resolution requirement. The example is shown in Table 1.

From the calculation, we establish new magnification buttons based on the down sampling factor. When users push the button, the platform will calculate appropriate virtual magnification and display parts of WSI in the viewer accordingly. Sample of 100x (low) and 400x (high) magnification displayed at 1200-by-900 pixel screen is shown in Fig. 2.

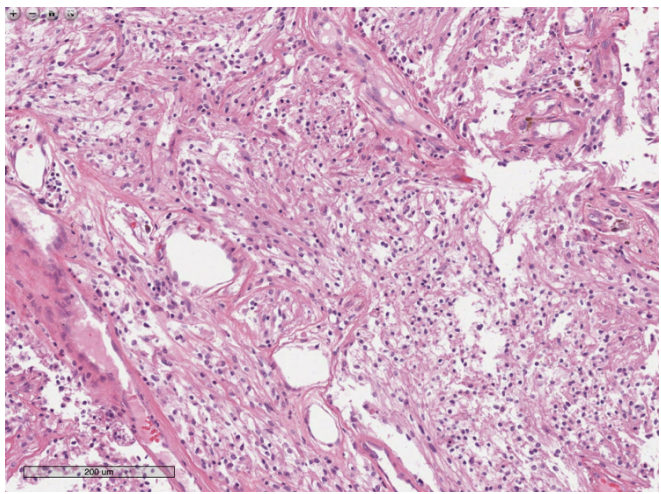

(a)

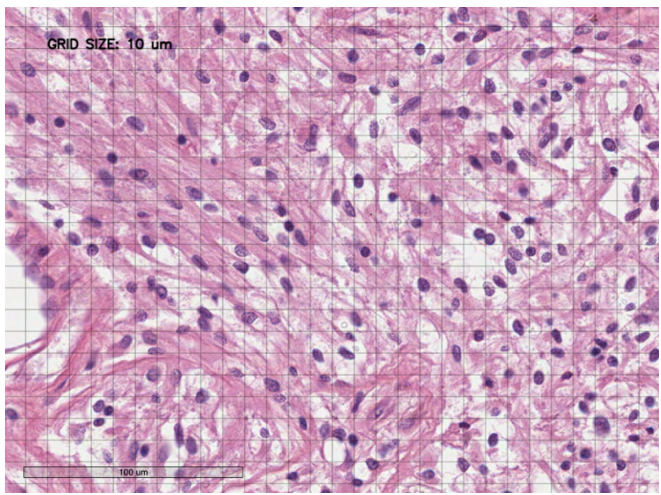

(b)

Sample of (a) low and (b) high magnification in OpenHI2 viewer displayed at 1200-by-900 pixel screen which correspond to 100x and 400x magnification in a typical optical microscope.

WSI loading time is also an issue which may reduce viewing and annotating performance. OpenHI2 utilizes an in-house efficient large image reading and writing library called libMI which could yield better image loading performance than the previous version while maintaining all functionalities.

Other small change about the viewer is that we added more visual ques to remind pathologists of the current magnification status. To do so, we added (a) adaptive scale bar at the bottom left of the viewer and (b) grid with specified grid size. Noted that we only display grid in higher magnification as can be seen in Fig. 2(b).

*2) Macro and micro image annotation*

Establishing large-scale annotation requires does not only requires the annotation to be precise and highly granularized. In other words, annotating the entire whole-slide image may be unnecessary since the slide may vastly contain redundant visual information. Because if this reason, pathologists routinely examine only a few diagnostic regions before they could come to a diagnosis for each slide. Meanwhile, to tackle the disagreements between annotators, annotations by different annotators should be made on the same diagnostic region, but they should not see each other's annotation.

With the mentioned scenario, OpenHI2 proposed a new module called to be annotated list. Each item on the list represents one diagnostic region selected by one of the annotators. The list is updated immediately once regions are added or removed can be seen by all annotators, while annotations on each diagnostic region are not. The process of selecting diagnostic regions correspond to macro scale annotation that is to find appropriate area in WSIs to make the diagnosis. We would like to note that, in our annotation trail, experienced pathologists do not have any disagreements at this macro annotation. Thus, collaborative diagnostic region selection which is not a blind process should meet the requirement.

*3) Inter-rater reliability tests*

To determine the concordance between multiple experts, qualitative methods that reflect the degree of agreement in annotations are needed. Developing the inter-rater reliability tests, we consider the annotation in two aspects: spatial agreements which reflect how much the two annotators utilize the visual information in the same area and grade agreements which tell how much they agree on the severity. The agreement can be calculated in three ways including percent agreement, Cohen's kappa, and F1-score. These conditions result in various APIs in OpenHI2 that the user could use, summarized in Table 2.

TABLE II.  OPENHI2 APIs FOR INTER-RATER RELIABILITY TESTS

| API | Description |
| --- | --- |
| compute_grade_score_overall() | Calculate the overall grading agreement. |
| compute_grade_score_overlapping() | Calculate the grading agreement only in the overlapping area. |
| compute_space_score_fscore() | Calculate F1-score based on selected regions regardless of grading. |
| compute_space_score_kappa() | Calculate kappa value based on selected regions regardless or grading. |

*4) Machine learning module*

One of the obstacles in the developments of statistical machine learning is data preparation, especially cleaning and organizing source data and labels. Instead of exporting annotations acquired by OpenHI and prepare the data for other frameworks, we provide a machine learning module built on PyTorch framework to directly develop new models, train, and predict on the platform via text-based commands. So far, we have successfully implement VGG, GoogLeNet, MobileNet, and Inception architectures.

### B. Dependencies

Extending from the original version [1], newly proposed modules in this paper are written in Python. libMI, written in C++ with Python wrapper, is used to support efficient image reading and writing. Flask is being used as a web server to support GUI web-based functions. PyTorch is used for the machine learning based analysis module. Documentations are made with Sphinx and available at https://bioai.gitlab.com/OpenHI. The platform also requires a MySQL server to store and organize loggings and annotators login information.

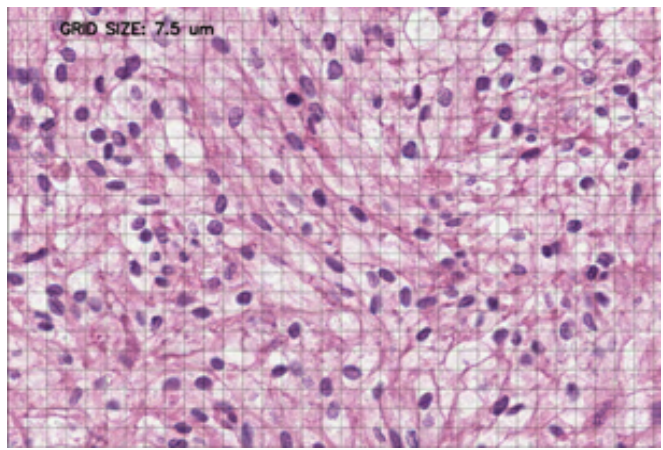
(a)

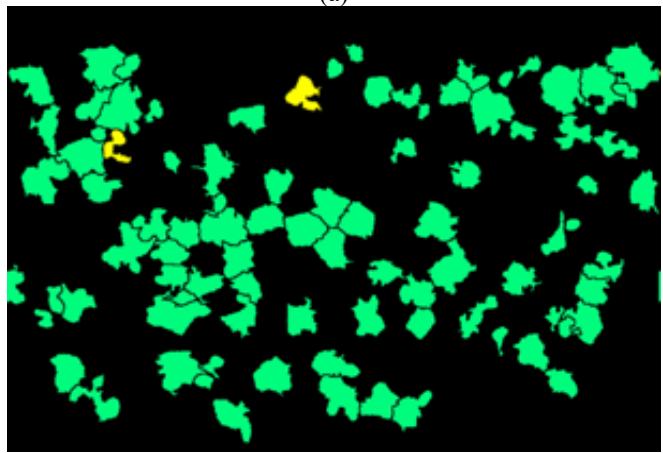
(b)

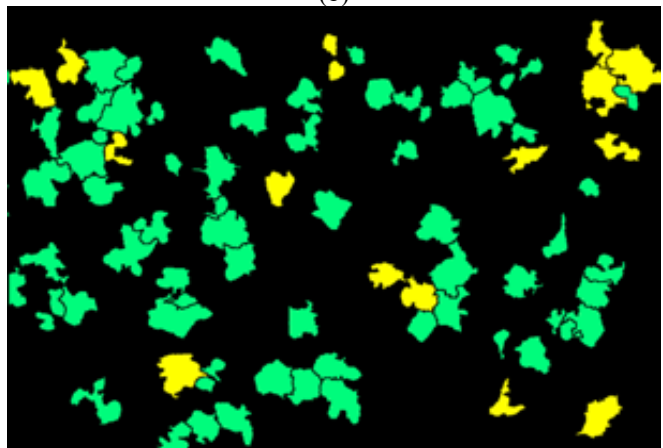
(c)

Sample of annotation from two pathologists (b-c) where (a) is the original image.

## IV. RESULTS

### A. Functionalities

*1) Pathological diagnosis routine support*

OpenHI2 has all basic functions to facilitate routine tasks of pathologists in making diagnosis based on digital slides including slide management, proper zooming and panning, and ability to display electronic clinical records. Furthermore, it is the first step as biomedical infrastructure platform toward digital pathology. OpenHI which adopts server-client scheme provide the flexibility for multiple users to access simultaneously access the same WSI file. Also, it requires no further installation on client machine since it can be accessed via any standardized web browsers. OpenHI can be deployed on a cloud infrastructure or servers in hospitals while requiring a LAMP or Docker environment which are widely available.

*2) Large-scale collaborative histopathological image annotation*

In OpenHI2, annotators can quickly zoom and pan to the diagnostic area with the to be annotated list, reducing time to seek for diagnostic region. Furthermore, different annotators can annotate the same diagnostic area since the to be annotated list is synced across all annotators. Combined with previously proposed cluster selection scheme, OpenHI2 can assist in creation of large-scale collaborative standardized annotation.

*B. Annotation trial*

We use OpenHI2 to perform an annotation trail. The annotation procedure consists of two steps: (1) diagnostic region selection and (2) cell cluster selection based on pre-segmented clusters. Two experienced pathologists participated in the trial. One WSI was used. In the first step 20 diagnostic regions were selected by one of the pathologists, then the two pathologists spend on average three non-continuous hours to finish the annotation with the total of 1,883 selected clusters by the two annotators. A sample diagnostic region with annotation from the two pathologists are shown in Fig. 3.

*C. Extendibility*

OpenHI2 is modularized and designed to be easy to extend with extensive documentation for each module. Currently, it can support all major WSI formats which is inherited from libMI integration. In clinical information, OpenHI2 supports XML file in the format from Genomic Data Commons of the US National Cancer Institute. Our software is free and open-source, available at https://gitlab.com/BioAI/OpenHI.

V. DISCUSSIONS

Although OpenHI2 may have all necessary functions for pathologists to use and make diagnoses, they may need additional training on digital or virtual slides before the new routine with digital pathology can be successfully and safely implemented in real-world settings. Standards and validation studies regarding the use computer displays for diagnosis is highly needed since histopathological diagnoses are sensitive to both image size/magnification and color. While OpenHI2 provides machine learning module especially designed for histopathological image analysis, the validity of trained classifiers is still questionable and there is no universal validation method that would proof medical validity of the system. Further work in this area is also needed.

A biomedical informatic system similar to OpenHI in digital pathology could be PACS which is currently responsible for managing all imaging data in hospitals. However, there are no implementation that uses DICOM which, in principle, could support multi-scale image as a file format to support WSIs. Currently, OpenHI2 lacks the ability to interoperate with external standardized servers such as PACS. The ability to work with other standardized system should be needed in the future. In the present, OpenHI2 is not only a web-based system but also a histopathological image library. APIs provided in the documentation are general and highly flexible, thus, it can be modified to serve different needs.

In some cases, the pre-segmentation method did not meet the desired precision which is to be able to select cells or nuclei. New segmentation methods should be added to the platform.

VI. CONCLUSION

Constant development in histopathological technique is becoming increasingly important to accurately and quickly diagnose tissue slides with the growing number of aging populations. OpenHI platform can potentially facilitate the transition from conventional to digital pathology. While it is functionally capable of performing necessary tasks, more trial runs and validations are still underway. Once pathology is digitized, it may be able to benefit from statistical machine learning techniques. The proposed platform also packs an annotation module which expert pathologists could use to precisely annotate the tissue and the annotation data can later be used to directly train artificial neural networks on the OpenHI2 platform without reorganizing the training data. This may result in a more rapid histopathological image classifiers development. The granular annotations acquired by the platform could also be used to precisely extract visual patterns which correspond to other medical information such as genotype, -omics, or clinical records.

ACKNOWLEDGMENT

This work has been supported by the National Key Research and Development Program of China [2018YFC0910404]; National Natural Science Foundation of China [61772409]; the consulting research project of the Chinese Academy of Engineering (The Online and Offline Mixed Educational Service System for "The Belt and Road" Training in MOOC China); Project of China Knowledge Centre for Engineering Science and Technology; the Innovation Team from the Ministry of Education [IRT\_17R86]; the Innovative Research Group of the National Natural Science Foundation of China [61721002]; and Professor Chen Li' s Recruitment Program for Young Professionals of "The Thousand Talents Plan".